\documentclass[conference]{IEEEtran}
\IEEEoverridecommandlockouts
\usepackage[compress,space]{cite}
\usepackage{amsmath,amssymb,amsfonts}
\usepackage{algorithmic}
\usepackage{graphicx}
\usepackage{textcomp}
\usepackage{xcolor}

\usepackage{multirow}
\def\BibTeX{{\rm B\kern-.05em{\sc i\kern-.025em b}\kern-.08em
    T\kern-.1667em\lower.7ex\hbox{E}\kern-.125emX}}
\usepackage{array}
\newcolumntype{L}{>{\arraybackslash}m{2cm}}

\begin{document}

\title{Speaker verification-derived loss and\\ data augmentation for DNN-based\\ multispeaker speech synthesis}

\author{\IEEEauthorblockN{Beáta Lőrincz$^{1,2}$, Adriana Stan$^1$, Mircea Giurgiu$^1$}
\IEEEauthorblockA{$^1$Communications Department,
Technical University of Cluj-Napoca, Romania \\
$^2$Computer Science Department, Babeș-Bolyai University of Cluj-Napoca, Romania\\
\{beata.lorincz, adriana.stan, mircea.giurgiu\}@com.utcluj.ro}
}

\maketitle

\begin{abstract}
Building multispeaker neural network-based text-to-speech synthesis systems
commonly relies on the availability of large amounts of high quality recordings from each speaker and conditioning the training process on the speaker's identity or on a learned representation of it. 
However, when little data is available from each speaker, or the number of speakers is limited, the multispeaker TTS can be hard to train and will result in poor speaker similarity and naturalness.

In order to address this issue, we explore two directions: forcing the network to learn a better speaker identity representation by appending an additional loss term; 
and augmenting the input data pertaining to each speaker using waveform manipulation methods.  
We show that both methods are efficient when evaluated with both objective and subjective measures. The additional loss term aids the speaker similarity, while the data augmentation improves the intelligibility of the multispeaker TTS system.
\end{abstract}


\section{Introduction}

Following the evolution of fundamental deep neural networks (DNN), the speech synthesis domain achieved results that were previously deemed unimaginable. Text-to-speech (TTS) implementations using recurrent architectures~\cite{wang2017tacotron, shen2018natural}, convolutional components~\cite{tachibana2018efficiently, ping2018deep}, transformers~\cite{ren2019fastspeech} or, more recently, the normalizing flow-based concept~\cite{valle2020flowtron} are currently driving not only the research community, but are also employed within large scale commercial systems.  These systems are capable of synthesising speech within the single or multiple speaker identity frameworks. 

Multispeaker TTS systems have the advantage of incorporating multiple vocal identities into a single network, and allow the user to select any of these identities without the need to change the model. As a result, the majority of the state-of-the-art synthesis systems also incorporate a multispeaker option. For instance, multispeaker capabilities were appended into two follow-up versions~\cite{gibiansky2017deep, ping2018deep} of the Deep Voice system~\cite{arik2017deep}. In~\cite{wang2018style} Tacotron~\cite{wang2017tacotron} was extended with a bank of embeddings which learn the speaker identities in an unsupervised manner. MultiSpeech~\cite{chen2020multispeech} is a transformer based multispeaker TTS system derived from~\cite{li2019neural}. \cite{park2019multi} extends ClariNet~\cite{ping2018clarinet} to support multispeaker speech synthesis. 


The most common approach for integrating multiple speakers in TTS systems is implemented by implicitly or explicitly learning a set of speaker-dependent representations, also called \emph{embeddings}. 
In~\cite{cooper2020zero} the authors compare various neural speaker embeddings and report that learnable dictionary encodings improve the speaker similarity for zero-shot speaker adaptation. \cite{liu2019cross} presents speaker, language and stress/tone embeddings used for TTS that can synthesize speech in multiple speaker identities and languages. Other techniques employ transfer learning methods to benefit from other speech processing applications, such as speaker verification~\cite{jia2018transfer} or speech recognition~\cite{inoue2020semi}. 

Yet, building multispeaker models requires large amounts of high quality speech recordings. 
When data is not readily available, data augmentation methods have been applied to overcome this barrier. In~\cite{hwang2020mel} augmentation methods such as frequency warping, duration and loudness control are used for producing additional speaker data. In~\cite{huybrechts2020low}, TTS systems are trained with data obtained through voice conversion mechanisms. Artificial speaker data and low-quality data usage is experimented within~\cite{Cooper2020} and the authors report improvements on the synthetic speech's naturalness.

One way to alleviate the task of learning speaker-dependent embeddings, as well as the lack of sufficient speaker data is to use a speaker verification (SV) network.
\cite{cho2020learning} examines whether the multispeaker TTS system can improve learning embeddings for the task of speaker verification, while~\cite{jia2018transfer} uses the speaker verification system to learn speaker embeddings used to condition the synthesised speech for different speakers. 
In~\cite{cai2020speaker} the authors present a model that uses a speaker verification component to enforce the learning of speaker identities. 

Considering the different methods of encoding speaker identities in multispeaker TTS, in this work we propose to enhance the learning of vocal identities by using speaker verification derived metrics and speaker augmented data within the training process.
Our work was developed in parallel and shares similarities with~\cite{cai2020speaker} in the sense that the feedback of the verification network is incorporated into the model within the loss function. However, the differences are twofold: firstly, we optimize both with cosine similarity and equal error rates calculated by the speaker verification model; secondly, we apply speaker augmentation methods and measure the effect of adding artificially produced data to the training dataset.

\section{Method overview}
\label{sec:method}

\begin{figure}[b]
    \centerline{\includegraphics[width=0.6\columnwidth, trim=0pt 0pt 0pt 0pt, clip]{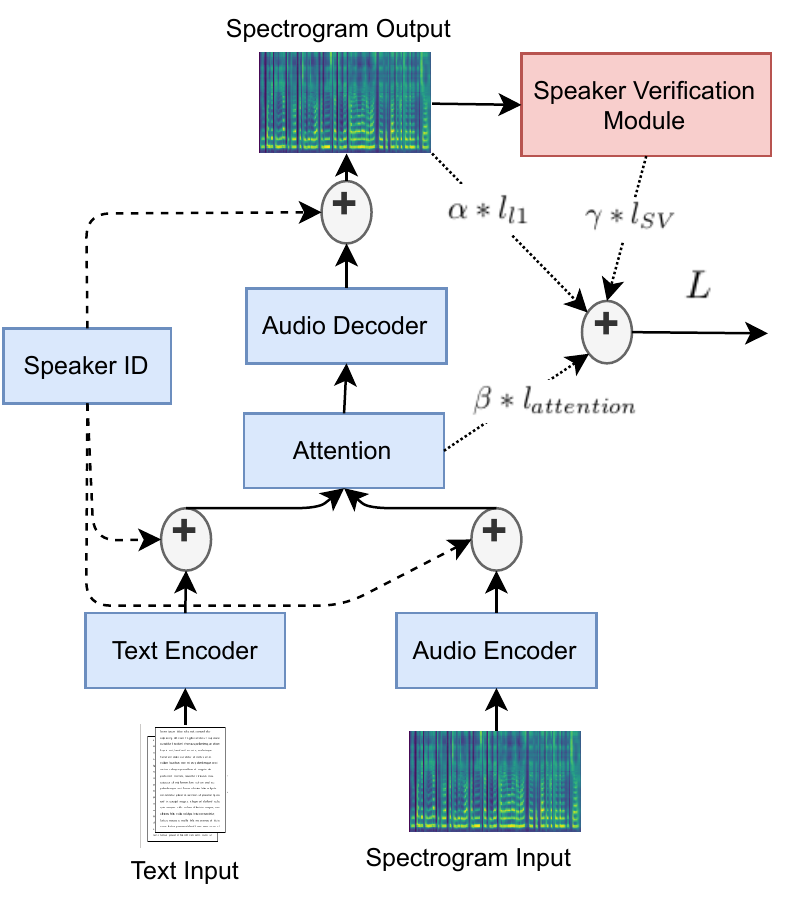}}
    \caption{The model architecture: speaker IDs are appended to several components of the TTS system; the speaker verification network delivers the objective measures used in the additional loss term.}
    \label{fig:architecture}
    \vspace{-10mm}
\end{figure}

The main objective of our study is to improve the speaker identity learning process in multispeaker TTS systems when a limited amount of speaker data is available.  
In order to achieve this, two scenarios are explored: (i) adding an additional loss term obtained from a pre-trained speaker verification network, and (ii) augmenting the data pertaining to each speaker through waveform manipulation methods.  
The starting point for all the experiments is a text-to-speech synthesis system based on a convolutional architecture with guided attention~\cite{tachibana2018efficiently}. The original model does not include speaker identity conditioning, and therefore its architecture was modified to incorporate a learned speaker embedding representation across all of its information channels (see Figure~\ref{fig:architecture}).
As a preliminary step for all the speaker similarity improvement methods, the original architecture, not conditioned on the speaker identity, was pre-trained with all the available data. The pre-training provides a stable uniform initialisation for the succeeding experiments.  

In the first speaker similarity improvement scenario we rely on the information provided by a pre-trained speaker verification (SV) system. The speaker embeddings generated by the SV network from the synthesised output and the embeddings extracted from natural samples are evaluated using the \emph{cosine similarity (CS)} function or the \emph{equal error rate (EER)} measure.
The CS value is equal to the batch average of the cosine similarity of the synthesised output with respect to its corresponding natural sample. For EER, 
depending on the training batch's speaker identities content, 
a list of random natural utterances are selected and compared with the synthesised output.
These speaker verification-based metrics are sequentially appended as additional weighted terms to the overall loss function of the TTS system, and referred to as $l_{SV}$. The resulting loss function of the TTS system is presented in Eq.~\ref{eq:loss}. $\alpha,\beta$ and $\gamma$ were empirically determined during the experiments. An overview of the loss modification process is shown in Figure~\ref{fig:architecture}.

\begin{equation}
\label{eq:loss}
    L =\alpha l_{l1} + \beta l_{attention} +\gamma l_{SV}
\end{equation}

As a result, we obtain a first set of systems for comparison: the baseline system conditioned on the speaker id (\texttt{B}); the baseline system trained with an additional loss term calculated with cosine similarity for the speaker embedding (\texttt{B+CS}); and the baseline system trained with an additional loss term calculated from the EER of the SV network (\texttt{B+E}). 

The other improvement scenario relies on the use of basic waveform manipulation methods to augment the data available for each speaker in the training set. The first method is inspired by~\cite{Cooper2020} and uses basic waveform resampling to modify the overall duration of the utterance. In the second method, we employed the Pitch Synchronous Overlap and Add (PSOLA)~\cite{moulines1990pitch} algorithm--commonly used in unit selection concatenative speech synthesis--to manipulate the duration and pitch of each of the speakers' utterances.
As opposed to the simple waveform resampling, PSOLA takes into account the pitch periods and thus yields, to a certain extent, a more natural output. 
By adding $l_{SV}$ to the TTS system, our main goal was to shift the learning process more towards the speaker identity, even if this meant that it could potentially alter the intelligibility or naturalness of the synthesised speech.
The data augmentation methods were combined with the previous loss term computation strategies and yielded additional systems' comparisons.





\section{Evaluation}
\label{sec:eval}

\subsection{Training data and speaker data augmentation}

The training data for our systems consists of the SWARA Romanian multispeaker parallel corpus~\cite{stan2017swara}. It includes 18 speakers: 10 female and 8 male voices, with the number of utterances per speaker being between 1000 and 1500. The data is sampled at 48kHz with 16 bps and was resampled at 16kHz to make the training process faster. 
The entire SWARA corpus was used to train the speaker verification network on which the additional loss terms are computed. 

For the multispeaker TTS systems' training we first selected 3 subsets from the corpus corresponding to various amounts of data for each speaker: \texttt{ALL}, \texttt{RND1} and \texttt{RND1-100}. A detailed description of these subsets is shown in Table~\ref{table:dataselection}. In the smaller subsets, parallel utterances were selected for all speakers such that their content does not influence the training and evaluation.

\begin{figure}[t]
    \centerline{\includegraphics[width=0.8\columnwidth, trim=0pt 0pt 0pt 0pt, clip]{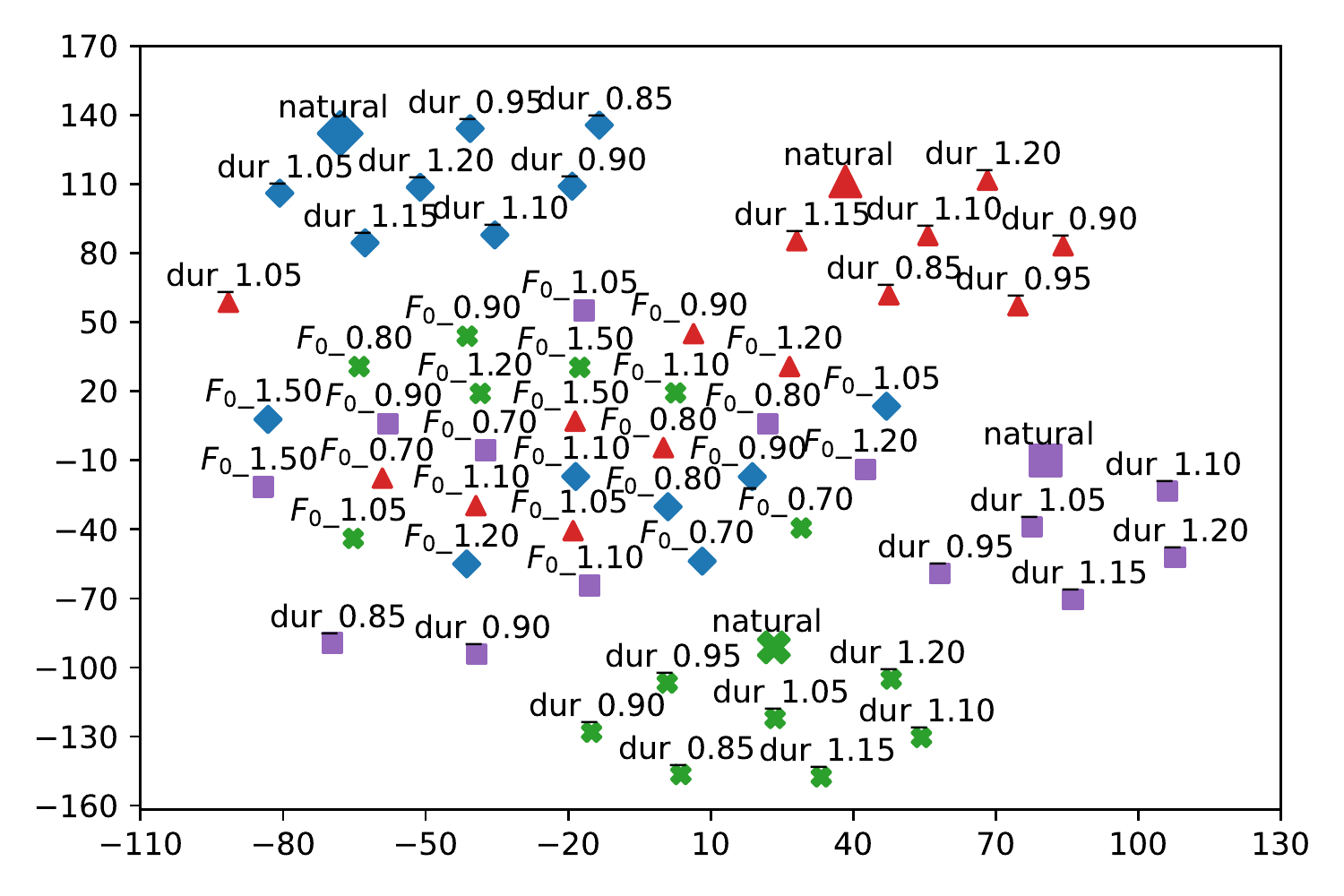}}
    \caption{t-SNE plots for speaker embeddings of natural and their corresponding augmented samples. The speakers are color coded, the original natural samples are marked as \emph{natural}, while the augmented versions are marked with \emph{dur} and \emph{$F_0$} followed by the modification ratio.}
    \label{fig:psola_dur_key}
    \vspace{-5mm}
\end{figure}

With respect to the data augmentation process, the waveform resampling was performed with the SoX\footnote{http://sox.sourceforge.net/} tool, as in~\cite{Cooper2020}. The different resampling ratios are shown in Table~\ref{table:dataselection}, and the system is referred to as \texttt{RND1-100-UP-DOWN}. 
The PSOLA-based waveform manipulation used the Pytsmod tool.\footnote{https://pypi.org/project/pytsmod/} Three separate modification sets were created: one through duration changes (\texttt{RND1-100-PSOLA-DUR}), another one by warping the $F_0$ values (\texttt{RND1-100-PSOLA-F0}), and the last one by modifying both the duration and the $F_0$ values (\texttt{RND1-100-PSOLA-MIX}).  
Because there is no guarantee that the PSOLA modifications do not alter the speaker identity, we used a speaker verification network trained on the VoxCeleb2 dataset~\cite{chung2018voxceleb2} to extract the speaker embedding vectors from all the natural and modified speech samples. A t-Distributed Stochastic Neighbour Embedding (t-SNE)~\cite{maaten2008visualizing} visualisation for a subset of them is shown in Figure~\ref{fig:psola_dur_key}. It can be noticed that the samples with the modified duration are closer to the natural ones, as opposed to the samples where the $F_0$ is modified. Using the t-SNE representations, in all three modification sets mentioned above, for each natural sample we selected the 4 closest modified samples in terms of Euclidean distance. This ensured that the number of samples available for each speaker in the \texttt{RND1-100} scenario is equal to the one in the \texttt{RND1} scenario, and their results can be directly compared.

\begin{table}[t!]
\renewcommand{\arraystretch}{1}
\caption{Training data selection and augmentation overview}
\label{table:dataselection}
\centering
\scriptsize
\begin{tabular}{p{0.2\columnwidth}  | p{0.6\columnwidth}}
\hline
\textbf{Data subset ID}  & \textbf{Description} \\
\hline
\textbf{ALL} (21h:54m) & 1000-1500 utterances selected from each speaker \\
\hline
\textbf{RND1} (10h:45m) & 500 utterances selected from each speaker \\ 
\hline
\textbf{RND1-100} (2h:24m) & 100 utterances selected from each speaker \\
\hline
\textbf{RND1-100-UP-DOWN}  (12h:01m) & 100 utterances selected from each speaker; each utterance resampled at 0.95, 0.975, 1.025 and 1.05 ratios, resulting in 500 utterances per speaker\\
\hline
\textbf{RND1-100-PSOLA-DUR}  (12h:35m) & 100 utterances selected from each speaker; for each utterance from a number of 7 time domain augmented files (with ratios 0.85, 0.90, 0.95, 1.05, 1.10, 1.15, 1.20) the best 4 selected, resulting in 500 utterances per speaker \\
\hline
\textbf{RND1-100-PSOLA-F0} (12h:31m) & 100 utterances selected from each speaker; for each utterance from a number of 7 frequency domain augmented files (with ratios 0.70, 0.80, 0.90, 1.05, 1.10, 1.20, 1.50) the best 4 selected, resulting in 500 utterances per speaker \\
\hline
\textbf{RND1-100-PSOLA-MIX} (12h:49m) & 100 utterances selected from each speaker; each utterances' duration modified by 1.3 and 0.8, and $F_0$ by 0.8, 1.2, resulting in 500 utterances per speaker \\
\hline
\end{tabular}
\end{table}

\subsection{Objective evaluation}

The starting point TTS system is based on a single speaker implementation\footnote{https://github.com/tugstugi/pytorch-dc-tts} of~\cite{tachibana2018efficiently} extended to a multispeaker scenario following ideas from a different implementation of the same tool.\footnote{https://github.com/CSTR-Edinburgh/ophelia} 
The TTS network without conditioning on the speaker identity was first pre-trained over 600 epochs. Using the learned weights for initialisation, all the evaluated synthesis systems (\texttt{B}, \texttt{B+CS}, \texttt{B+E}) with the various data selection criteria were trained up to 1000 epochs. One exception was the \texttt{RND1-100} system trained over 5000 epochs so as to address the fewer training samples.

The speaker verification network followed the implementation of~\cite{chung2020defence},\footnote{https://github.com/clovaai/voxceleb\_trainer} used only the SWARA corpus for training and was appended to the TTS system architecture (see Figure~\ref{fig:architecture}).

The output of the multispeaker TTS systems was evaluated both objectively, in terms of word error rate (WER) and equal error rate (EER), as well as subjectively through listening tests.\footnote{
Audio samples for the presented systems and samples generated with data augmentation methods are available at: https://speech.utcluj.ro/multispeaker\_tts/} The WER should estimate the intelligibility of the system and was computed using a general purpose high-quality automatic speech recognition (ASR) system in Romanian~\cite{8906555}. The ASR's WER over the natural samples was 7.43\%. Given a well trained SV network, the EER should be able to measure the objective speaker similarity of the synthetic output with respect to the natural samples. Therefore we used a speaker verification network trained on a large corpora--not including SWARA--and published by the authors of~\cite{chung2020defence}. This external SV network ensured that there is no bias or overfitting of the network for the speakers available in the training data. 
The SV network evaluation over natural samples from each speaker resulted in an EER value of 4\%.

\begin{table}[t!]
\renewcommand{\arraystretch}{1}
\caption{\textbf{WER} and \textbf{EER} (\%) results for the evaluated systems}
\label{table:eer_wer_results}
\centering
\begin{tabular}{|L|ccc|ccc|} \hline

\multirow{2}{*}{\textbf{}} &  \multicolumn{3}{c|}{\textbf{WER [\%]}}&  \multicolumn{3}{c|}{\textbf{EER [\%]}} \\ \cline{2-7}
              & \textbf{B} &\textbf{B+CS} & \textbf{B+E} & \textbf{B}& \textbf{B+CS}&\textbf{B+E}\\ \cline{1-7}
\textbf{ALL} & 9.54 & 7.66 & 8.26 & 6.94 & 4.66 & 4.66 \\\hline
\textbf{RND1} & 9.99 & 8.67 & 9.86 & 4.86 & 4.00 & 4.66 \\\hline
\textbf{RND1-100} & 11.13 & 10.21 & 13.26 & 5.55 & 5.33 & 5.33 \\\hline
\textbf{RND1-100-UP-DOWN} & 12.42 & - & - & 8.66 & - & - \\ \hline
\textbf{RND1-100-PSOLA-F0} & 14.04 & 15.75 & 14.18 & 8.66 & 10.66 & 11.33\\ \hline
\textbf{RND1-100-PSOLA-DUR} & 11.84 & 13.62 & 10.32 & 8.33 & 6.25 & 10.00 \\ \hline
\textbf{RND1-100-PSOLA-MIX} & 10.05 & - & 16.00 & 9.72 & - & 6.94 \\ \hline

\hline
\end{tabular}
\end{table}

\begin{figure*}[th!]
\renewcommand{\arraystretch}{1.0}
\renewcommand{\tabcolsep}{0.1cm}
    \centerline{\includegraphics[width=2\columnwidth, trim=0pt 0pt 0pt 0pt, clip]{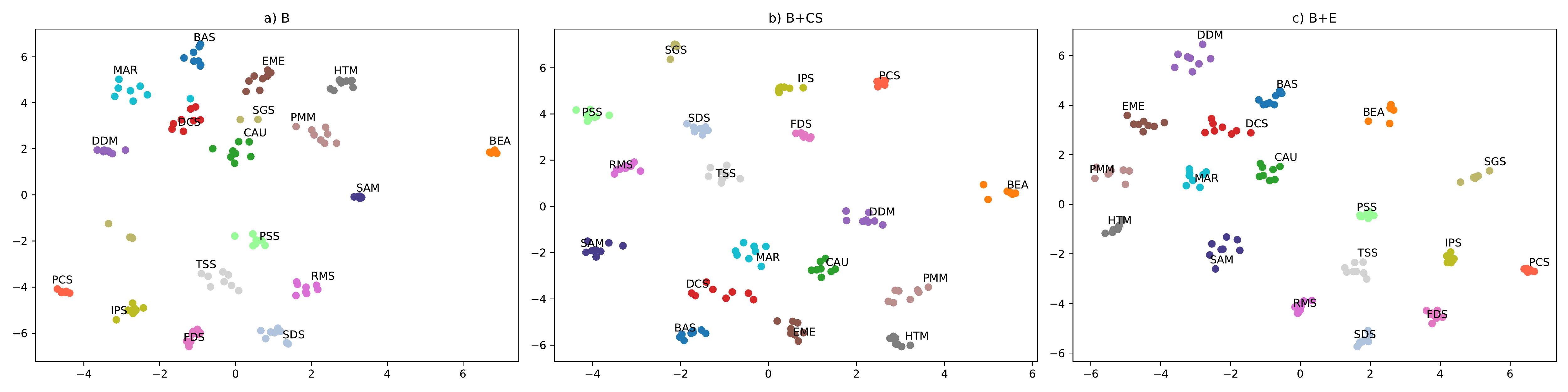}}
    \caption{t-SNE plots of the speaker embeddings extracted by the evaluation speaker verification network from the synthesised outputs of the \texttt{ALL} systems in the three training scenarios: \texttt{B}, \texttt{B+CS} and \texttt{B+E}. The speakers are color coded.}
    \label{fig:tSNE_systems}
    \vspace{-5mm}
\end{figure*}

\begin{figure}[t]
    \centerline{\includegraphics[width=\columnwidth, trim=0pt 0pt 0pt 0pt, clip]{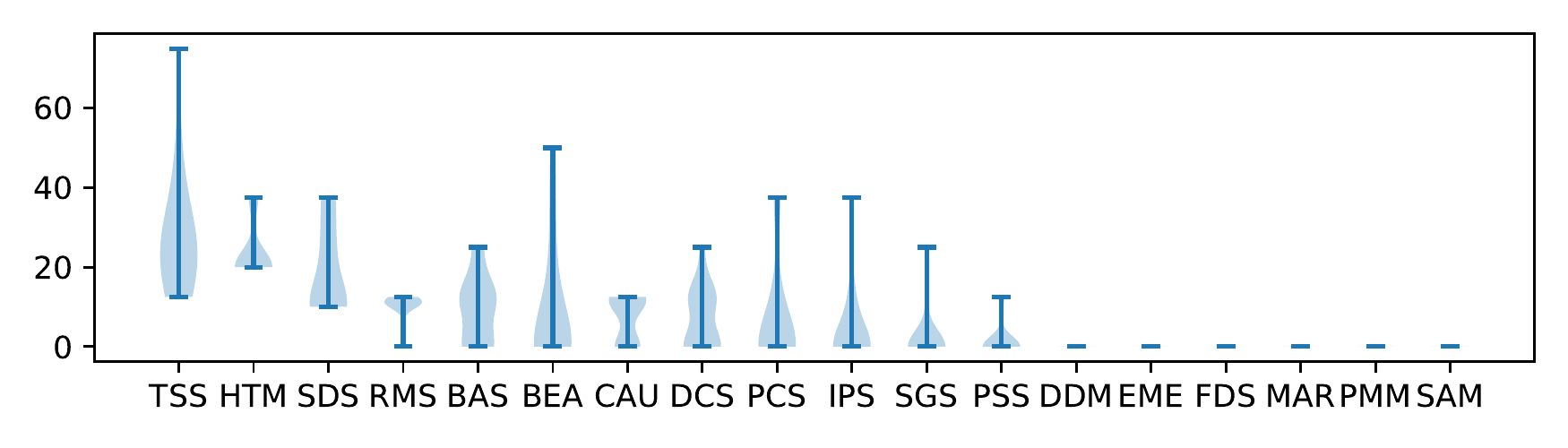}}
    \caption{Violin plots of the EER values obtained by each speaker for all the evaluated TTS systems.}
    \label{fig:violin_speaker_eer}
    \vspace{-5mm}
\end{figure}


From each multispeaker TTS system 8 samples were generated for each of the 18 speakers available in SWARA, using the Griffin-Lim vocoder~\cite{griffin1984signal}. The lexical content was the same across all systems and speakers. The resulting 144 samples were then transcribed with the Romanian ASR tool and the WER was computed against the ground truth transcripts. In the EER evaluation, the same 144 synthesised utterances were paired with natural samples pertaining to the same or to a random different speaker identity resulting in 288 pairs. The WER and EER values for the evaluated systems are summarized in Table~\ref{table:eer_wer_results}. Given the higher values of the baseline systems for the \texttt{RND1-100-UP-DOWN} and \texttt{RND1-100-PSOLA-MIX} datasets, we did not train and evaluate the \texttt{B+CS} or \texttt{B+E} systems, hence the missing results in Table~\ref{table:eer_wer_results}.

In terms of objective intelligibility, appending the SV-based loss term, on average, lowers the WER for the systems trained using only the natural speech samples (i.e. \texttt{ALL}, \texttt{RND1} and \texttt{RND1-100}). Better results are obtained when using the cosine similarity measure as opposed to EER. It may be the case that the EER-based loss term is not linearly dependent on the spectrogram output, and therefore does not translate into a well-behaved gradient.
On the other hand, the data augmentation seems to affect the intelligibility of the TTS system. The worst effect over the intelligibility is the result of the $F_0$ manipulation (\texttt{RND1-100-PSOLA-F0} - WER: 14.04\%). However, it is interesting to notice that when selecting a mix of duration and $F_0$ manipulated samples (\texttt{RND1-100-PSOLA-MIX}) the WER (10.05\%) is very close to the one obtained when a similar number of natural samples is used for training (\texttt{RND1} - WER: 9.99\%). This means, in this case, that a similar intelligibility can be obtained by using only a fifth of the data pertaining to each speaker, and augmenting it with PSOLA-derived samples. 
With respect to the objective speaker similarity measure, again, the additional loss term (\texttt{B+CS} and \texttt{B+E}) improves the EER for the natural sample-based systems, with the best performing system being the one which uses the cosine similarity term. However, the training data waveform manipulations does not contribute to a better learning of the speaker identity. This might be explained by the fact that the augmented samples introduce a larger speaker variability and the identity of the speakers in the TTS output is affected by the changes in the duration or $F_0$ values. 

To understand the above results better, we plotted the t-SNE representations of the speaker embeddings extracted by the evaluation SV network for the systems trained with all the available speaker data (\texttt{ALL}) while adding each additional loss term (\texttt{CS} or \texttt{E}). The results are shown in  
Figure~\ref{fig:tSNE_systems}. We can observe that in the case of the baseline system, speakers \emph{SGS} and \emph{MAR} have samples that are further apart in the embedding space, and the distance between these is smaller in case of the \texttt{B+CS} and \texttt{B+E} systems. While for the \emph{DDM} speaker, the additional terms spread the embeddings through the space. This means that the behaviour of the TTS network favours some of the speakers, while neglecting or averaging the other's characteristics. Further on, we looked at the average EERs for each speaker in the training set over all systems and plotted the violin representations of these values in Figure~\ref{fig:violin_speaker_eer}. As it can be easily noticed, there is a clear difference between the learned identities across the speakers. For example \emph{TSS} and \emph{HTM} commonly achieve EERs above 25\%, while speakers \emph{SAM}, \emph{PMM} or \emph{MAR} have EERs closer to 0\%. We should mention the fact that within this ordering or the t-SNE plots above, there is no definite distinction between the male and female speakers. However this type of ordering and visualisation could provide a more reliable means to select a suitable subset of speaker identities for a multispeaker TTS system.

\subsection{Subjective evaluation}

\begin{figure}[b!]
    \centerline{\includegraphics[width=0.9\columnwidth, trim=0pt 0pt 0pt 0pt, clip]{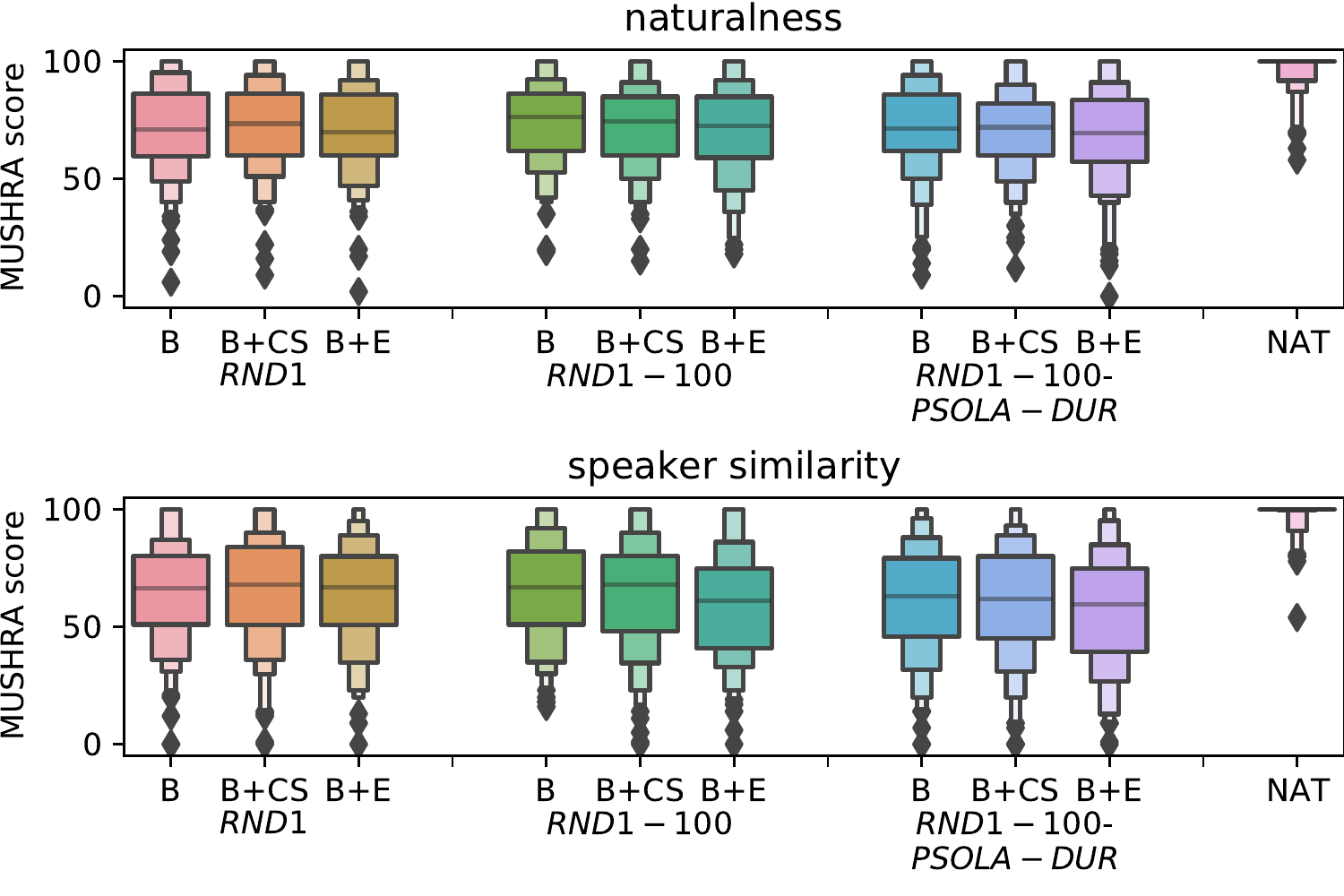}}
    \caption{Letter-value plots of the listening test MuSHRA scores in the naturalness and speaker similarity sections.}
    \label{fig:ls_boxenplot}
    \vspace{-5mm}
\end{figure}

As the objective evaluation of the TTS systems cannot truly reflect the human perception, we also employed a listening test evaluation for a subset of the trained systems.
A MUlti Stimulus test with Hidden Reference and Anchor (MuSHRA) method\footnote{ITU-R Recommendation BS.1534-1} was used to collect the subjective evaluations from 27 native listeners. No lower anchor was used in the test set. Naturalness and speaker similarity results are shown in Figure~\ref{fig:ls_boxenplot}. It can be noticed that although the differences are rather small, the ordering of the systems and improvement methods obtained in the objective evaluation holds true within the listening test as well. This is one additional result supporting the use of high-quality ASR and SV systems in the evaluation of TTS systems' output.


\section{Conclusions}
\label{sec:conslusions}
In this paper we proposed two methods to improve the learning of speaker identities in multispeaker text-to-speech systems. The first one was based on using additional loss terms based on the embedding obtained by a speaker verification network trained on the same data as the TTS system. The second one employed waveform level manipulation methods to augment the available data pertaining to each speaker within the training set. Objective and subjective results showed that the data augmentation can in some cases reduce the required amount of training data to a fifth of the original size, and that the cosine similarity measure increases the intelligibility and speaker similarity of the synthetic output irrespective of the amount of training data available for each speaker. One additional result was the different performance of the TTS system and the improvement methods across the speakers. This leaves space for exploring speaker-dependent training strategies in multispeaker TTS systems.

\section*{Acknowledgment}
\small
This work was funded through a grant from the Romanian Ministry of Research and Innovation, PCCDI – UEFISCDI, project number PN-III-P1-1.2-PCCDI-2017-0818/73. We would also like to thank the listening test volunteers for their evaluation.

\bibliographystyle{IEEEtran}
\footnotesize
\bibliography{master}

\begin{thebibliography}{10}
\providecommand{\url}[1]{#1}
\csname url@samestyle\endcsname
\providecommand{\newblock}{\relax}
\providecommand{\bibinfo}[2]{#2}
\providecommand{\BIBentrySTDinterwordspacing}{\spaceskip=0pt\relax}
\providecommand{\BIBentryALTinterwordstretchfactor}{4}
\providecommand{\BIBentryALTinterwordspacing}{\spaceskip=\fontdimen2\font plus
\BIBentryALTinterwordstretchfactor\fontdimen3\font minus
  \fontdimen4\font\relax}
\providecommand{\BIBforeignlanguage}[2]{{%
\expandafter\ifx\csname l@#1\endcsname\relax
\typeout{** WARNING: IEEEtran.bst: No hyphenation pattern has been}%
\typeout{** loaded for the language `#1'. Using the pattern for}%
\typeout{** the default language instead.}%
\else
\language=\csname l@#1\endcsname
\fi
#2}}
\providecommand{\BIBdecl}{\relax}
\BIBdecl

\bibitem{wang2017tacotron}
Y.~Wang, R.~Skerry-Ryan, D.~Stanton, Y.~Wu, R.~J. Weiss, N.~Jaitly, Z.~Yang,
  Y.~Xiao, Z.~Chen, S.~Bengio \emph{et~al.}, ``Tacotron: Towards end-to-end
  speech synthesis,'' \emph{arXiv preprint arXiv:1703.10135}, 2017.

\bibitem{shen2018natural}
J.~Shen, R.~Pang, R.~J. Weiss, M.~Schuster, N.~Jaitly, Z.~Yang, Z.~Chen,
  Y.~Zhang, Y.~Wang, R.~Skerrv-Ryan \emph{et~al.}, ``{Natural TTS Synthesis by
  Conditioning WaveNet on Mel Spectrogram Predictions},'' in \emph{2018 IEEE
  International Conference on Acoustics, Speech and Signal Processing
  (ICASSP)}.\hskip 1em plus 0.5em minus 0.4em\relax IEEE, 2018, pp. 4779--4783.

\bibitem{tachibana2018efficiently}
H.~Tachibana, K.~Uenoyama, and S.~Aihara, ``Efficiently trainable
  text-to-speech system based on deep convolutional networks with guided
  attention,'' in \emph{2018 IEEE International Conference on Acoustics, Speech
  and Signal Processing (ICASSP)}.\hskip 1em plus 0.5em minus 0.4em\relax IEEE,
  2018, pp. 4784--4788.

\bibitem{ping2018deep}
W.~Ping, K.~Peng, A.~Gibiansky, S.~O. Arik, A.~Kannan, S.~Narang, J.~Raiman,
  and J.~Miller, ``Deep voice 3: 2000-speaker neural text-to-speech,''
  \emph{Proc. ICLR}, pp. 214--217, 2018.

\bibitem{ren2019fastspeech}
Y.~Ren, Y.~Ruan, X.~Tan, T.~Qin, S.~Zhao, Z.~Zhao, and T.-Y. Liu, ``Fastspeech:
  Fast, robust and controllable text to speech,'' in \emph{Advances in Neural
  Information Processing Systems}, 2019, pp. 3171--3180.

\bibitem{valle2020flowtron}
R.~Valle, K.~Shih, R.~Prenger, and B.~Catanzaro, ``Flowtron: an autoregressive
  flow-based generative network for text-to-speech synthesis,'' \emph{arXiv
  preprint arXiv:2005.05957}, 2020.

\bibitem{gibiansky2017deep}
A.~Gibiansky, S.~Arik, G.~Diamos, J.~Miller, K.~Peng, W.~Ping, J.~Raiman, and
  Y.~Zhou, ``Deep voice 2: Multi-speaker neural text-to-speech,'' in
  \emph{Advances in neural information processing systems}, 2017, pp.
  2962--2970.

\bibitem{arik2017deep}
S.~O. Arik, M.~Chrzanowski, A.~Coates, G.~Diamos, A.~Gibiansky, Y.~Kang, X.~Li,
  J.~Miller, A.~Ng, J.~Raiman, S.~Sengupta, and M.~Shoeybi, ``Deep voice:
  Real-time neural text-to-speech,'' \emph{arXiv preprint arXiv:1702.07825},
  2017.

\bibitem{wang2018style}
Y.~Wang, D.~Stanton, Y.~Zhang, R.~Skerry-Ryan, E.~Battenberg, J.~Shor, Y.~Xiao,
  F.~Ren, Y.~Jia, and R.~A. Saurous, ``Style tokens: Unsupervised style
  modeling, control and transfer in end-to-end speech synthesis,'' \emph{arXiv
  preprint arXiv:1803.09017}, 2018.

\bibitem{chen2020multispeech}
M.~Chen, X.~Tan, Y.~Ren, J.~Xu, H.~Sun, S.~Zhao, and T.~Qin, ``Multispeech:
  Multi-speaker text to speech with transformer,'' \emph{arXiv preprint
  arXiv:2006.04664}, 2020.

\bibitem{li2019neural}
N.~Li, S.~Liu, Y.~Liu, S.~Zhao, and M.~Liu, ``Neural speech synthesis with
  transformer network,'' in \emph{Proceedings of the AAAI Conference on
  Artificial Intelligence}, vol.~33, 2019, pp. 6706--6713.

\bibitem{park2019multi}
J.~Park, K.~Zhao, K.~Peng, and W.~Ping, ``Multi-speaker end-to-end speech
  synthesis,'' \emph{arXiv preprint arXiv:1907.04462}, 2019.

\bibitem{ping2018clarinet}
W.~Ping, K.~Peng, and J.~Chen, ``Clarinet: Parallel wave generation in
  end-to-end text-to-speech,'' \emph{arXiv preprint arXiv:1807.07281}, 2018.

\bibitem{cooper2020zero}
E.~Cooper, C.-I. Lai, Y.~Yasuda, F.~Fang, X.~Wang, N.~Chen, and J.~Yamagishi,
  ``Zero-shot multi-speaker text-to-speech with state-of-the-art neural speaker
  embeddings,'' in \emph{ICASSP 2020-2020 IEEE International Conference on
  Acoustics, Speech and Signal Processing (ICASSP)}.\hskip 1em plus 0.5em minus
  0.4em\relax IEEE, 2020, pp. 6184--6188.

\bibitem{liu2019cross}
Z.~Liu and B.~Mak, ``Cross-lingual multi-speaker text-to-speech synthesis for
  voice cloning without using parallel corpus for unseen speakers,''
  \emph{arXiv preprint arXiv:1911.11601}, 2019.

\bibitem{jia2018transfer}
Y.~Jia, Y.~Zhang, R.~Weiss, Q.~Wang, J.~Shen, F.~Ren, P.~Nguyen, R.~Pang,
  I.~Lopez~Moreno, Y.~Wu \emph{et~al.}, ``Transfer learning from speaker
  verification to multispeaker text-to-speech synthesis,'' \emph{Advances in
  neural information processing systems}, vol.~31, pp. 4480--4490, 2018.

\bibitem{inoue2020semi}
K.~Inoue, S.~Hara, M.~Abe, T.~Hayashi, R.~Yamamoto, and S.~Watanabe,
  ``Semi-supervised speaker adaptation for end-to-end speech synthesis with
  pretrained models,'' in \emph{ICASSP 2020-2020 IEEE International Conference
  on Acoustics, Speech and Signal Processing (ICASSP)}.\hskip 1em plus 0.5em
  minus 0.4em\relax IEEE, 2020, pp. 7634--7638.

\bibitem{hwang2020mel}
Y.~Hwang, H.~Cho, H.~Yang, D.-O. Won, I.~Oh, and S.-W. Lee, ``Mel-spectrogram
  augmentation for sequence to sequence voice conversion,'' \emph{arXiv
  preprint arXiv:2001.01401}, 2020.

\bibitem{huybrechts2020low}
G.~Huybrechts, T.~Merritt, G.~Comini, B.~Perz, R.~Shah, and J.~Lorenzo-Trueba,
  ``Low-resource expressive text-to-speech using data augmentation,''
  \emph{arXiv preprint arXiv:2011.05707}, 2020.

\bibitem{Cooper2020}
E.~Cooper, C.-I. Lai, Y.~Yasuda, and J.~Yamagishi, ``{Can Speaker Augmentation
  Improve Multi-Speaker End-to-End TTS?}'' in \emph{Proc. Interspeech 2020},
  2020, pp. 3979--3983.

\bibitem{cho2020learning}
J.~Cho, P.~Zelasko, J.~Villalba, S.~Watanabe, and N.~Dehak, ``Learning speaker
  embedding from text-to-speech,'' \emph{arXiv preprint arXiv:2010.11221},
  2020.

\bibitem{cai2020speaker}
Z.~Cai, C.~Zhang, and M.~Li, ``From speaker verification to multispeaker speech
  synthesis, deep transfer with feedback constraint,'' \emph{arXiv preprint
  arXiv:2005.04587}, 2020.

\bibitem{moulines1990pitch}
E.~Moulines and F.~Charpentier, ``Pitch-synchronous waveform processing
  techniques for text-to-speech synthesis using diphones,'' \emph{Speech
  communication}, vol.~9, no. 5-6, pp. 453--467, 1990.

\bibitem{stan2017swara}
A.~Stan, F.~Dinescu, C.~{\c{T}}iple, {\c{S}}.~Meza, B.~Orza, M.~Chiril{\u{a}},
  and M.~Giurgiu, ``{The SWARA speech corpus: A large parallel Romanian read
  speech dataset},'' in \emph{2017 International Conference on Speech
  Technology and Human-Computer Dialogue (SpeD)}.\hskip 1em plus 0.5em minus
  0.4em\relax IEEE, 2017, pp. 1--6.

\bibitem{chung2018voxceleb2}
J.~S. Chung, A.~Nagrani, and A.~Zisserman, ``Voxceleb2: Deep speaker
  recognition,'' \emph{arXiv preprint arXiv:1806.05622}, 2018.

\bibitem{maaten2008visualizing}
L.~v.~d. Maaten and G.~Hinton, ``Visualizing data using t-sne,'' \emph{Journal
  of machine learning research}, vol.~9, no. Nov, pp. 2579--2605, 2008.

\bibitem{chung2020defence}
J.~S. Chung, J.~Huh, S.~Mun, M.~Lee, H.~S. Heo, S.~Choe, C.~Ham, S.~Jung, B.-J.
  Lee, and I.~Han, ``In defence of metric learning for speaker recognition,''
  \emph{arXiv preprint arXiv:2003.11982}, 2020.

\bibitem{8906555}
A.~{Georgescu}, H.~{Cucu}, and C.~{Burileanu}, ``{Kaldi-based DNN Architectures
  for Speech Recognition in Romanian},'' in \emph{2019 International Conference
  on Speech Technology and Human-Computer Dialogue (SpeD)}, 2019, pp. 1--6.

\bibitem{griffin1984signal}
D.~Griffin and J.~Lim, ``{Signal estimation from modified short-time Fourier
  transform},'' \emph{IEEE Transactions on Acoustics, Speech, and Signal
  Processing}, vol.~32, no.~2, pp. 236--243, 1984.

\end{thebibliography}

\end{document}